# A comparison of the carrier density at the surface of quantum wells for different crystal orientations of silicon, gallium arsenide and indium arsenide

Ryan Hatcher and Chris Bowen

Samsung Advanced Logic Lab, 12100 Samsung Blvd Austin, Tx 78754

We report the carrier densities at the surface of single-crystal quantum wells as a function of material, orientation and well width. We include wells constructed from silicon, gallium arsenide and indium arsenide with three crystal orientations, (100), (110) and (111), included for each material. We find that the $\Delta_2$ states in a silicon (100) quantum well have the smallest density near the surface of the slab. Inspection of the planar average of the carrier densities reveals a characteristic shape that depends on the material and orientation, which leads to a varying degree of suppression or enhancement of the density near the surface. The physics responsible for the suppression or enhancement of the density near the surface can be traced to a constraint imposed by the symmetry of quantum well wavefunction on the phases of the bulk Bloch states of the crystal from which it can be constructed.

The performance of scaled electronics is increasingly dependent on surface interactions as technology nodes evolve to ever more confined geometries such as Ultra Thin Body-FETs[1], FinFETs[2,3], and nanowires[4,5]. Furthermore, non-silicon channel materials such as SiGe, Ge, InGaAs etc… are being considered for future technology nodes[6,7]. Here we compare the relative electron carrier density outside an idealized quantum well structure for different materials and orientations as a function of the well width. The goal is to identify trends in the surface density as a function of the material and crystal orientation that can be traced to an effect that depends on the symmetry of the underlying bulk Bloch states[8]. To that end, an artificial passivation scheme is employed in order to isolate the contribution to the surface density from the bulk symmetry properties of a given material and crystal orientation.

Calculations were performed with the Socorro[9] code using norm-conserving pseudopotentials[10], a plane wave basis set, and the generalized-gradient approximation for exchange and correlation[11]. The plane wave cutoff energy is 544 eV and the two dimensional Brillouin zone was sampled using the Monkhorst-Pack technique[12] with an irreducible set of k-points corresponding to a 9x9x1 grid.

Examples of quantum wells (QWs) for silicon and III-V crystals are shown in Figure 1a and the minimal supercells are shown for the (100), (110) and (111) orientations in Figure 1b. The QWs vary in thickness from 3-5nm with approximately 2nm of vacuum. The structures are constructed by cleaving a perfect crystal and passivating in order to eliminate dangling bonds and charge polarization between surfaces. For QWs with nonpolar surfaces, each dangling bond at the surface is passivated with one hydrogen atom. For QWs constructed from either GaAs or InAs with surfaces in either the (100) or (111) orientations, the surfaces are polar and passivated with artificial hydrogen-like atoms. Because a cation in a III-V zinc blende crystal provides $\frac{3}{4}$ of an electron per bond, each dangling bond associated with a cation is passivated with an artificial hydrogen-like atom of charge $\frac{5}{4}$. Similarly, each anion in a III-V zinc blende crystal contributes $\frac{5}{4}$ of an electron per bond and therefore each anion dangling bond is passivated with a hydrogen-like atom with charge $\frac{3}{4}$. The use of unrelaxed structures with artificial hydrogen-like atoms is perhaps unsettling but necessary in order to provide a comparison of the inversion layer densities at the surface of each QW not biased by material- and orientation-dependent surface reconstructions. The goal instead is to isolate the effects of bulk symmetry properties for a given crystal orientation on the structure of the electronic states that make up the inversion layer and in turn how this structure affects the density at and beyond the surface.

For (100) QWs with polar surfaces and an even number of atomic planes, one surface is cation-terminated and the other anion-terminated. For (100) polar surfaces with an odd number of atomic planes two calculations are performed, one with both surfaces anion-terminated anion and another with both surfaces cation-terminated. There are two inequivalent (111) surfaces that alternate – one with three dangling bonds per terminal atom and one with only one dangling bond per terminal atom. For the sake of simplicity, we only consider QWs with an even number of atomic planes chosen such that each terminating layers have only one dangling bond per atom. For III-V QWs with (111) orientation, one surface is cation-terminated and the other surface is anion-terminated as shown in Figure 1b.

In Figure 2, the planar averages of the conduction band minimum (CBM) electron densities are shown for nine QWs. For supercells shown in Figure 1, confinement leads to a two dimensional Brillouin zone (BZ) parallel to the surface of the QW. The III-V QWs each have a non-degenerate CBM at $\Gamma$ whereas the CBM for silicon has an orientation-dependent degeneracy stemming from the six-fold degeneracy of the bulk silicon CBM along the $\Delta$ lines to X. The CBM states for a silicon QW are found by projecting the bulk CBM states onto the 2D BZ of the QW for a given orientation. For each orientation of silicon in Figure 2, there is an inset with the bulk CBM lobes to illustrate how the states will map onto the 2D BZ. Note that any pair of states that have mirror symmetry in the direction of quantization will have their degeneracy broken by quantization. For a (100) silicon QW, there are four degenerate states, $\Delta_4$, that correspond to bulk CBM states along the $\Delta$ lines parallel to the surface of the QW. Quantization lifts the degeneracy between the remaining two states, $\Delta_2$, which have mirror symmetry across the plane at the center of the QW and correspond to the bulk CBM states that lie along $\Delta$ lines perpendicular to the surface of the QW. For a (110) silicon QW, there are two degenerate states entirely in the plane of the QW. The remaining four states have a component in the plane of the QW and another component perpendicular to the plane of the QW with mirror symmetry for pairs of states. These four states are two-fold degenerate with partial degeneracy lifted due to quantization. Lastly, the six CBM states for a silicon (111) QW each have a component in the plane of the QW and another component perpendicular to this plane but there is no mirror symmetry between the states. Therefore, the Si (111) CBM states retain six-fold degeneracy.

The distribution of the CBM densities in Figure 2 reveals an unintuitive effect first discussed by Boykin[13-16] and described for the case of the $\Delta_2$ and $\Delta_4$ states in a silicon (100) QW[8]. The surface flaring due to this parity dependent effect is especially pronounced in InAs where a node in the density nearly forms in the middle of the QW indicating a large odd component in the Bloch states of the bulk crystal. Because of this parity induced suppression of the density in the middle of the well for GaAs and InAs, the density is effectively pushed against the sides of the well resulting in substantially higher electron density at the surface compared to silicon. Conversely, the silicon QW wavefunctions are peaked near the center of the well indicating a relatively large even component of the underlying Bloch states of the bulk crystal leading to an effective suppression of the surface density. In addition, the Si(100) $\Delta_2$, Si(110) $\Delta_4$ and Si(111) CBM states correspond to bulk Bloch states with a component of the Bloch wave vector in the direction of quantization, which limits the spatial extent of the modulation of the density to roughly ${}^{1}/{}_{k_\perp}$ thereby further limiting the contribution to the surface density from the odd component of the bulk Bloch states.

In order to quantify the effect for comparison, we define the integrated surface density (ISD) as a figure of merit that corresponds to the fraction of the CBM density outside the surface of the well. Thus, the ISD is calculated by integrating the planar average of the CBM density from the last atomic plane in the QW to infinity as shown in Figure 3a and dividing by the total density of the CBM state then averaging over both surfaces. Thus, the ISD is a dimensionless quantity given as a percentage and corresponds to the fraction of each CBM electron just outside each surface of the well.

The ISDs for different materials and orientations for well widths of about 4nm are shown in Figure 3b. For the case of one of the $\Delta_2$ states in a silicon (100) QW, only 0.02% of each electron is outside the well, which is two orders of magnitude smaller than the ISDs of most of the III-V QWs regardless of orientation. The InAs (111) QW has the largest ISD at about 3.61%. Overall, the III-V QWs have larger ISDs than silicon QWs. The average ISD over non-degenerate states in silicon is 0.25%, which is a factor of 10 smaller than 2.57%, the average over all III-V states. Even the silicon (111) QW state, which at 0.40% is the silicon QW state with the largest ISD, is a factor of 3.8 *smaller* than the ISD of GaAs (110) with both surfaces Ga-terminated, which at 1.52% is the smallest among the III-V QWs.

The ISD dependencies on well width are shown in Figure 4 for QWs between 3 and 5nm. The most significant property affecting the ISD is the material with silicon leading to the smallest ISD and InAs the largest. The difference between silicon and the III-Vs is quite large. At 3nm, the average ISD for silicon QWs is 0.50%, a factor of 7.0 larger than 3.48%, which is the average for the III-V QWs. At 5nm, the silicon average is 0.15% whereas the III-V average is 1.99%, a factor of 13.3. Within the III-Vs, the GaAs QWs (solid lines) have ISDs that are systematically smaller than those of the InAs QWs (dotted lines) by a factor ranging from 1.3 for 3nm wells to 1.5 for 5nm wells for a given orientation and termination. The termination of (100) III-V QWs also has a significant effect on the ISD, with anion-terminated surfaces leading to surface densities about 1.7 times larger than cation-terminated surfaces for all well widths. This observation is consistent with an effect that has been reported in GaN/$Al_xGa_{1-x}N$ heterostructures[17]. We expect that this sensitivity to the terminating species depends on the chemistry of the passivation scheme. However, for any covalent passivation, an anion-terminated surface should be passivated with atoms that serve as cation and vice versa (e.g. a GaAs-InAs heterostructure). Therefore, we would expect this effect to be present in real III-V QWs with polar surfaces and covalent passivation. Finally, the orientation contributes to the relative surface densities. For all three materials, the (111) QWs have the largest or nearly the largest surface densities and the (100) surfaces have the smallest (cation-terminated for the III-Vs).

In conclusion, we have shown that the ISDs in GaAs and InAs QWs are significantly larger than for silicon QWs. We attribute the difference to the bulk properties of the crystals from which the QWs are constructed. In particular, the parity of the underlying bulk Bloch states lead to either a suppression or enhancement of the ISD for states with large even or odd components, respectively.

The authors acknowledge helpful conversations with Borna Obradovic and Mark Rodder.

Figure Captions

1 (a). Examples of Silicon and III-V Quantum with (100) surfaces. Note that for III-V QWs with polar surfaces, artificial hydrogen-like atoms are used to passivate the surface in order to eliminate both dangling bonds and charge accumulation without reconstruction.

1 (b). QW supercells shown for each orientation for silicon and III-Vs. Note that the (100) and (111) QW are shown with an even number of atomic planes, which results in one surface cation-terminated and the other surface anion-terminated. For III-V (100) QWs with an odd number of atomic planes, two calculations are performed as well – one with both surfaces cation-terminated and another with both surfaces anion-terminated.

2. The planar average of the density of the lowest conduction state(s) for quantum wells constructed from silicon, GaAs and InAs with three crystal orientations. For the III-V QWs, there is a non-degenerate conduction band minimum at G. For bulk silicon, there are six degenerate lowest conduction states, one along each D line to X as shown in the insets for different the different orientations of silicon. In a (100) silicon quantum well, four of these D states, referred to as the $D_4$ states, are in the plane of the QW and are degenerate. The remaining two D states, referred to as the $D_2$ states, lie along the direction of quantization, which breaks the degeneracy as indicated with the striped lobes in the insets. Similarly, for Si (110) QW's, the six-fold bulk degeneracy is partially lifted by quantization. The two states in the plane parallel to the surface are degenerate and referred to as the $D_2$ states, and the four remaining states each have a component in the direction of quantization which breaks the degeneracy for the states with reflection symmetry. Thus the $D_4$ states are two-fold degenerate.

3. (a) The planar averages of the three nearly degenerate conduction band minima for a Si (100) QW 40.7Å in diameter. The shaded region under each planar average in the inset is the ISD, which starts from the edge of the terminal silicon atom in the QW and extends to infinity. The ISD is the average over both surfaces and is given as a percentage of the total state density.

3 (b) The ISDs for the conduction band minima are given for Si, GaAs and InAs QWs of about 40Å thick oriented along the [100], [110] and [111] directions. In addition, two ISDs are given for the (100) InAs and GaAs QW with an odd number of atomic planes – one with both surfaces cation-terminated (i.e. Ga-Ga or In-In) and one with both surfaces anion-terminated (i.e. As-As). The (100) InAs and GaAs QW's with an even number of atomic planes include cation-terminated surface and one anion-terminated surface (i.e. Ga-As or In-As).

4. The ISDs as a function of well width from 30-50Å. Silicon QWs are indicated by dashed lines, GaAs solid lines and InAs dotted lines. The circles indicate (100) orientation, squares indicated (110) orientation and (111) QWs are labeled with triangles. The ISDs for III-V QW's increase monotonically as the well width decreases due to confinement effects.

Figure 1a

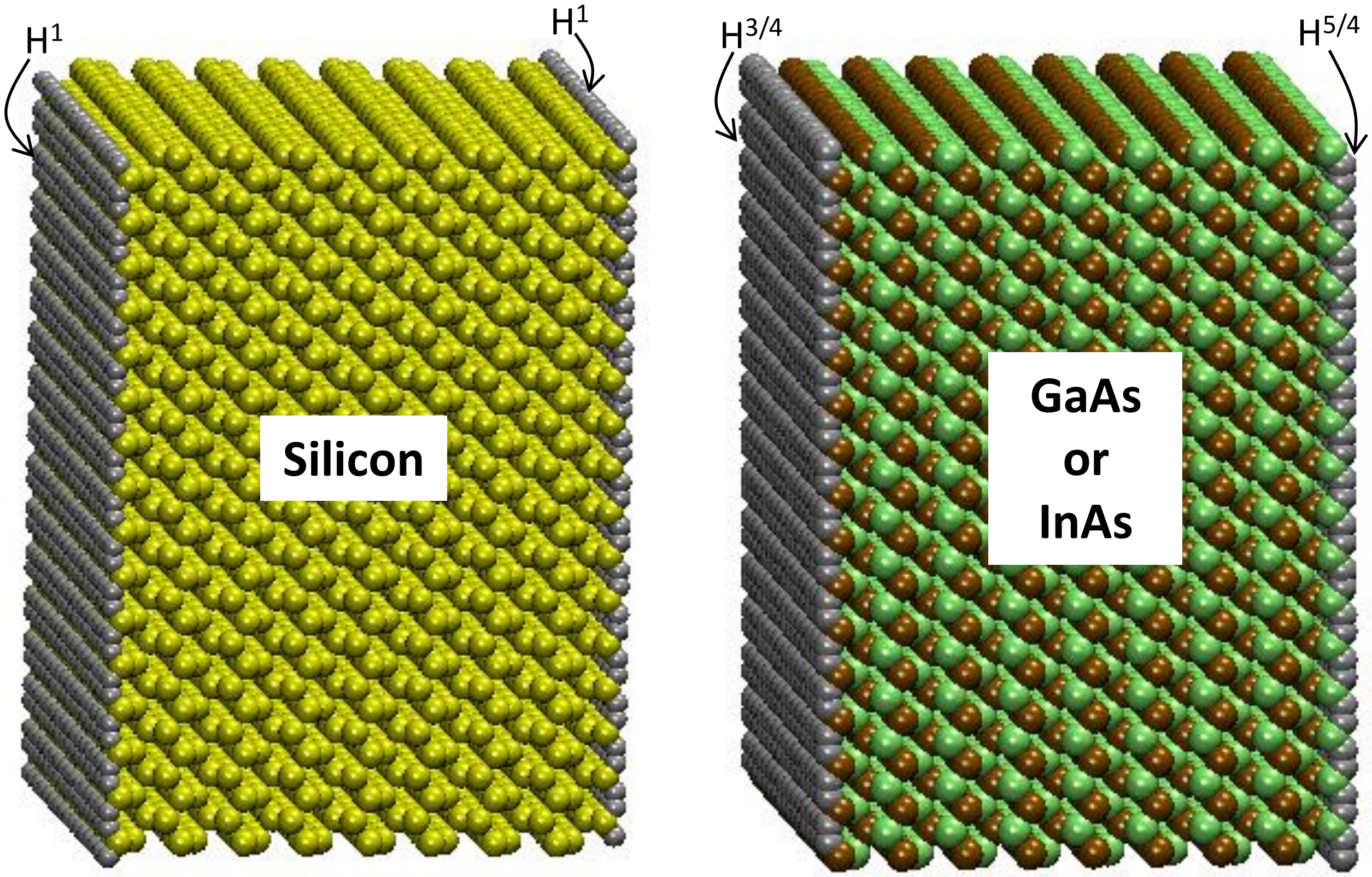

Figure 1a. Examples of Silicon and III-V Quantum with (100) surfaces. Note that for III-V QWs with polar surfaces, artificial hydrogen-like atoms are used to passivate the surface in order to eliminate both dangling bonds and charge accumulation without reconstruction.

Figure 1b

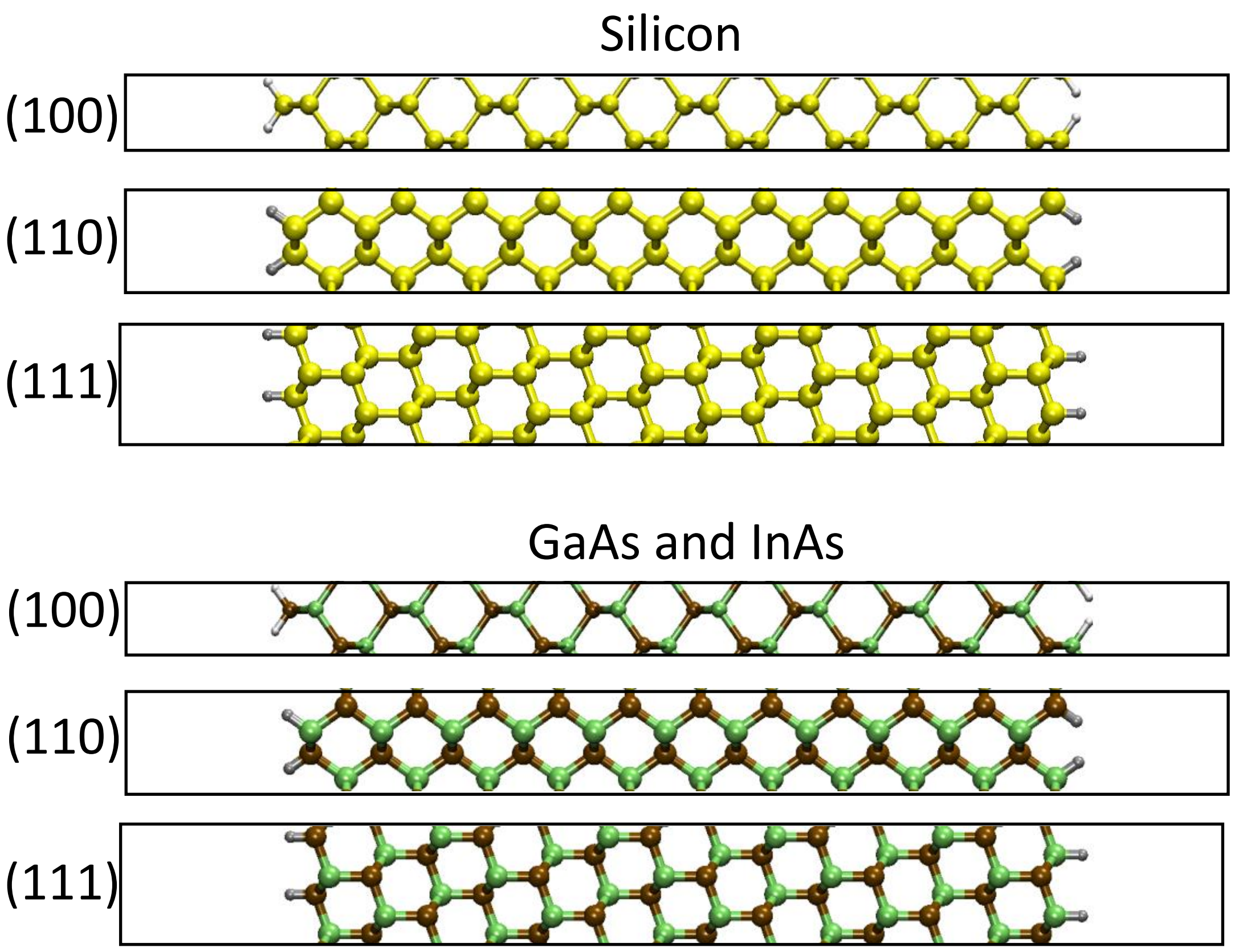

Figure 1b. QW supercells shown for each orientation for silicon and III-Vs. Note that the (100) and (111) QW are shown with an even number of atomic planes, which results in one surface cation-terminated and the other surface anion-terminated. For III-V (100) QWs with an odd number of atomic planes, two calculations are performed as well – one with both surfaces cation-terminated and another with both surfaces anion-terminated.

Figure 2

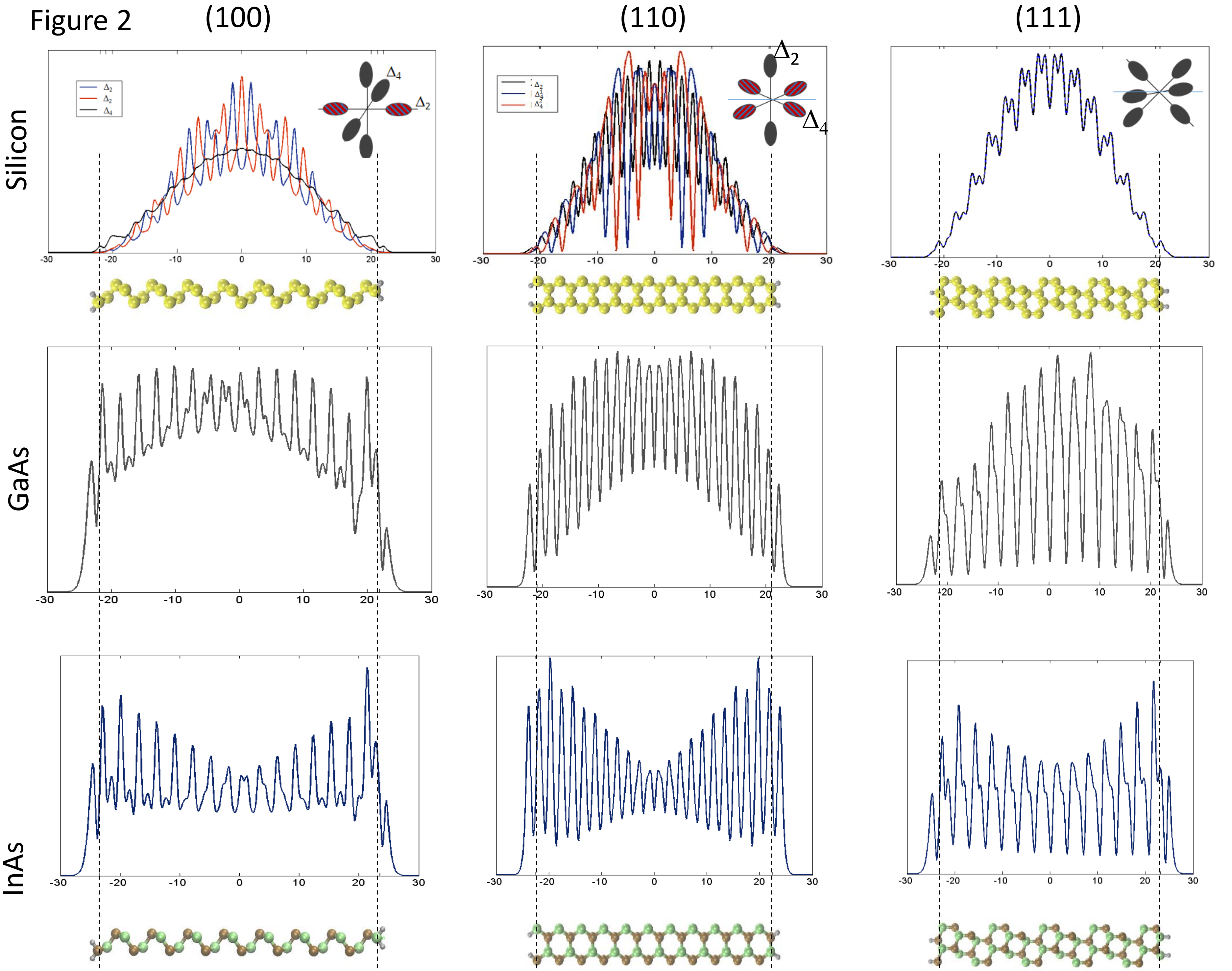

Figure 2. The planar average of the density of the lowest conduction state(s) for quantum wells constructed from silicon, GaAs and InAs with three crystal orientations. For the III-V QWs, there is a non-degenerate conduction band minimum at $\Gamma$. For bulk silicon, there are six degenerate lowest conduction states, one along each $\Delta$ line to X as shown in the insets for different the different orientations of silicon. In a (100) silicon quantum well, four of these $\Delta$ states, referred to as the $\Delta_4$ states, are in the plane of the QW and are degenerate. The remaining two $\Delta$ states, referred to as the $\Delta_2$ states, lie along the direction of quantization, which breaks the degeneracy as indicated with the striped lobes in the insets. Similarly, for Si (110) QW's, the six-fold bulk degeneracy is partially lifted by quantization. The two states in the plane parallel to the surface are degenerate and referred to as the $\Delta_2$ states, and the four remaining states each have a component in the direction of quantization which breaks the degeneracy for the states with reflection symmetry. Thus the $\Delta_4$ states are two-fold degenerate.

Figure 3a

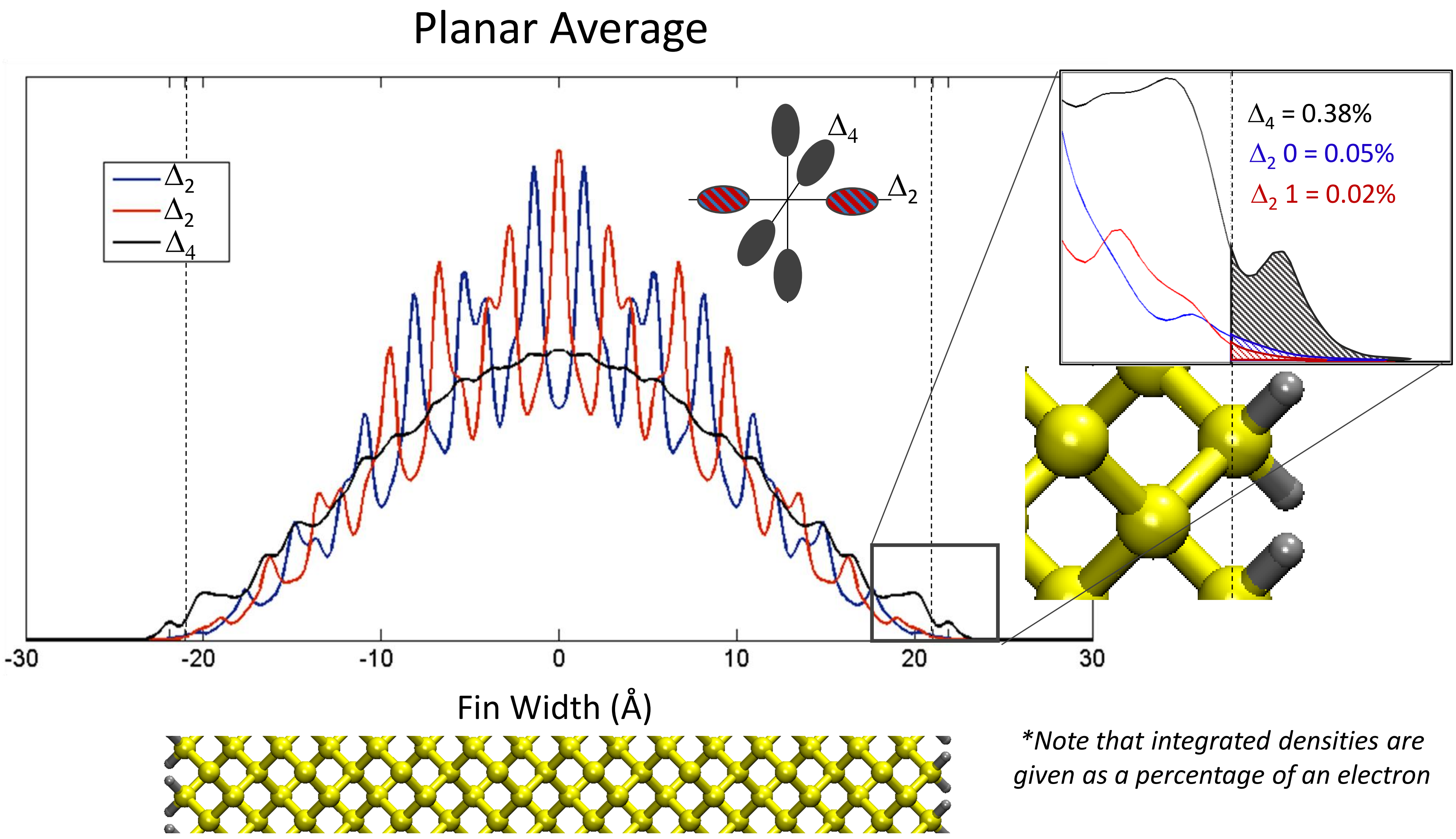

Figure 3a. The planar averages of the three nearly degenerate conduction band minima for a Si (100) QW 40.7Å in diameter. The shaded region under each planar average in the inset is the integrated surface density, which starts from the edge of the terminal silicon atom in the QW and extends to infinity. The integrated surface density is the average over both surfaces and is given as a percentage of the total state density.

# Figure 3b

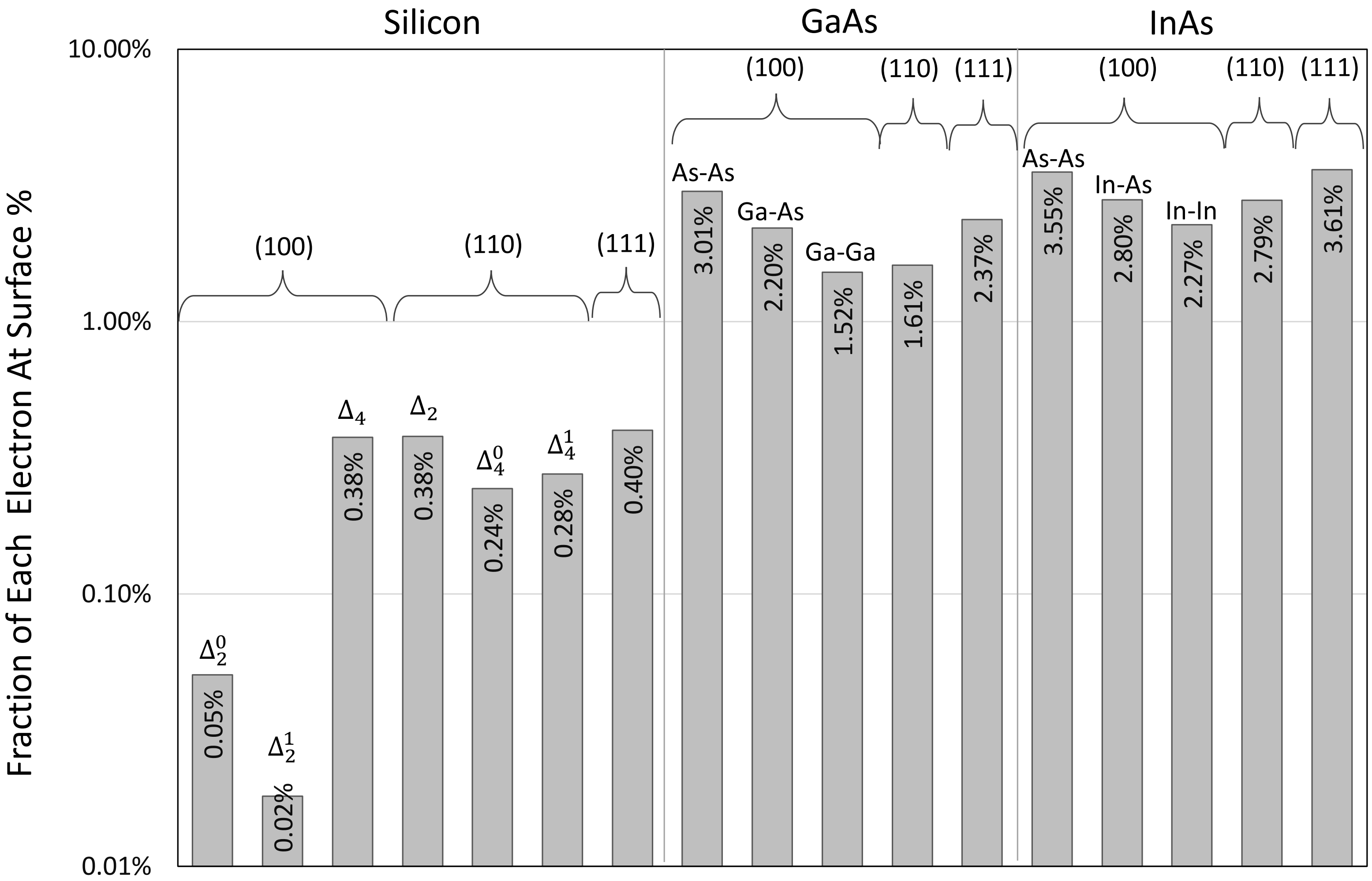

Figure 3b. The integrated surface densities for the conduction band minima are given for Si, GaAs and InAs QWs of about 40Å thick oriented along the [100], [110] and [111] directions. In addition, two integrated surface densities are given for the (100) InAs and GaAs QW with an odd number of atomic planes – one with both surfaces cation-terminated (i.e. Ga-Ga or In-In) and one with both surfaces anion-terminated (i.e. As-As). The (100) InAs and GaAs QW's with an even number of atomic planes include cation-terminated surface and one anion-terminated surface (i.e. Ga-As or In-As).

# Figure 4 Si, GaAs and InAs QWs

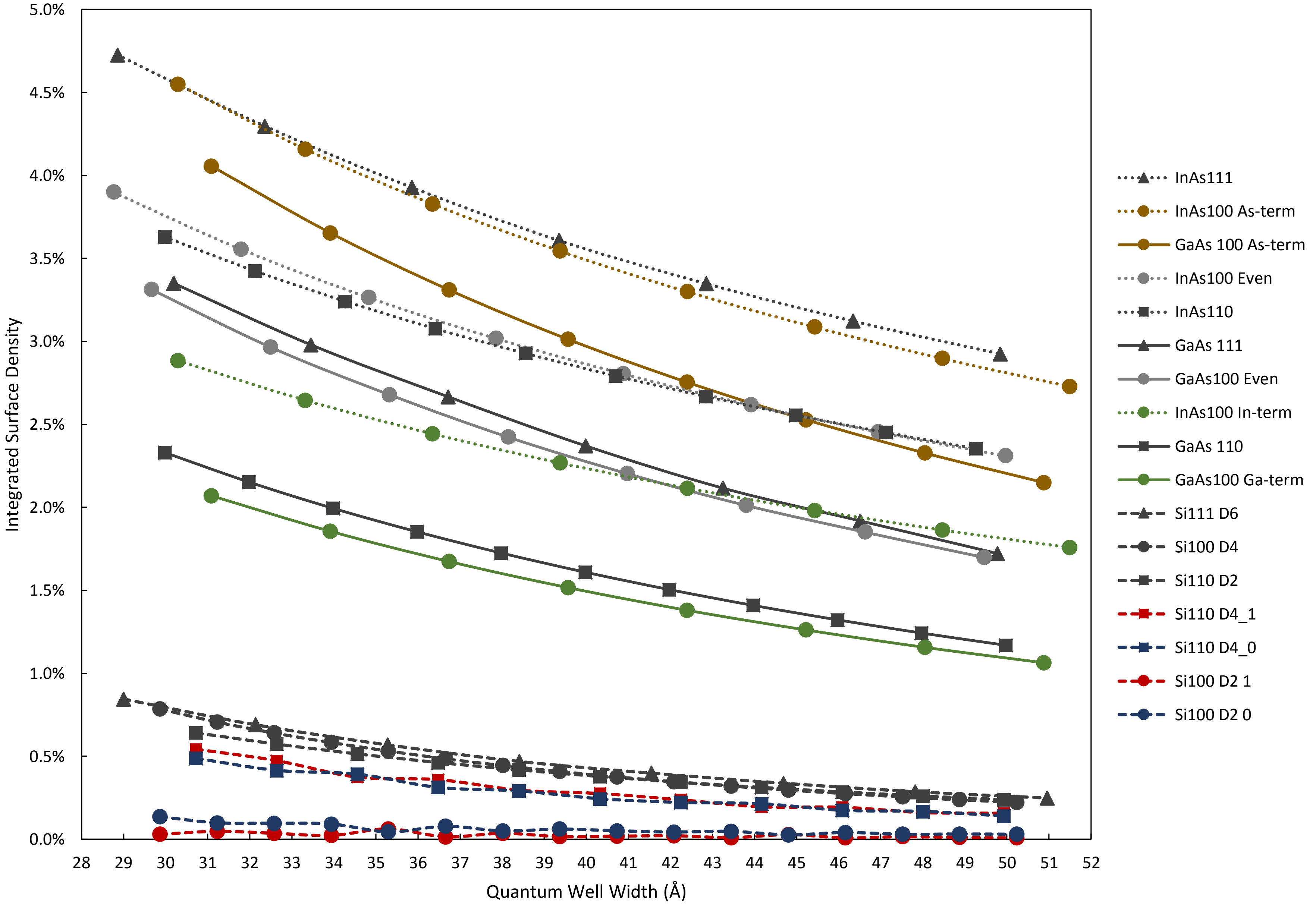

Figure 4a. The integrated surface densities as a function of well width from 30-50Å. Silicon QWs are indicated by dashed lines, GaAs solid lines and InAs dotted lines. The circles indicate (100) orientation, squares indicated (110) orientation and (111) QWs are labeled with triangles. The integrated probability densities for III-V QW's increase monotonically as the well width decreases due to confinement effects.